\documentclass[superscriptaddress,twocolumn,showpacs,amsmath,amssymb]{revtex4}

\usepackage{graphicx}
\usepackage{times}
 
\def\ii{{\mathrm{i}}}
\def\ee{{\mathrm{e}}}
\def\U#1{{%
\def\O{\mbox{O}}
\def\u{\mbox{u}}
\mathcode`\u=\mu
\mathcode`\O=\Omega
\mathrm{#1}}}
\def\sub#1{_{\rm #1}}
\def\ket#1{|\mbox{$#1$}\rangle}
\def\dd{{\mathrm{d}}}
\def\bracketi#1#2{\langle\mbox{$#1$}|\mbox{$#2$}\rangle}

\begin{document}

\title{
Two-photon interference of photon pairs
created in photonic crystal fibers}

\author{T. Nakanishi}
\affiliation{Department of Electronic Science and Engineering,
Kyoto University, Kyoto 615-8510, Japan}
\affiliation{CREST, Japan Science and Technology Agency, Tokyo 102-0075, Japan}
\author{K. Sakemi}
\affiliation{Department of Electronic Science and Engineering,
Kyoto University, Kyoto 615-8510, Japan}
\author{H. Kobayashi}
\affiliation{Department of Electronic Science and Engineering,
Kyoto University, Kyoto 615-8510, Japan}
\author{K. Sugiyama}
\affiliation{Department of Electronic Science and Engineering,
Kyoto University, Kyoto 615-8510, Japan}
\affiliation{CREST, Japan Science and Technology Agency, Tokyo 102-0075, Japan}
\author{M. Kitano}
\affiliation{Department of Electronic Science and Engineering,
Kyoto University, Kyoto 615-8510, Japan}
\affiliation{CREST, Japan Science and Technology Agency, Tokyo 102-0075, Japan}

\date{\today}

\email{t-naka@kuee.kyoto-u.ac.jp}

\begin{abstract}
We investigate a method to produce photon pairs by four-wave mixing in
photonic crystal fibers (PCFs).
By controlling the wavelength of the pump light,
which determines the phase matching condition for  four-wave mixing, 
we can obtain a broader spectrum of photon pairs 
than undesired Raman-scattered photons.
We observe two-photon interference of photon pairs from a PCF 
with the help of an unbalanced Mach-Zehnder interferometer.
Two-photon interference 
fringes with 83\% visibility, which exceeds the classical limit of 50\%,
are obtained.
\end{abstract}


\maketitle

\section{Introduction}

Photon pairs have been utilized in various fields 
such as fundamental physics and have found applications 
in quantum computation and communication,
because they show various types of quantum correlations,
for example, polarization correlation
\cite{PhysRevLett.49.1804,Kwiat:1995p147},
time-frequency correlation \cite{Franson:1989p152,Kwiat:1993},
and spatial correlation \cite{DAngelo:2004p17}.
Two-photon interference is one of the pure quantum phenomena
attributed to quantum correlations.
In experiments on two-photon interference,
each photon pair behaves like a quantum object
called a ``{\it biphoton},'' whose effective energy (or frequency)
is  twice that of the original photons,
and the interference fringe of the photon pair has half the period of
a one-photon interference fringe \cite{Edamatsu:2002p145}.
Two-photon interference can be applied to
high-resolution lithographic technology that overcomes the
 classical diffraction limit
\cite{Boto:2000p47}.

Three-wave mixing (TWM) in $\chi^{(2)}$ nonlinear crystals
has been extensively used  for the generation of  photon pairs
 \cite{Kwiat:1995p147,Kwiat:1999p148}.
Recently, four-wave mixing (FWM) in $\chi^{(3)}$ nonlinear material
has also been studied as an effective method to produce 
entangled photon pairs \cite{Edamatsu:2004p46,Fiorentino:2002p146,
Li:2005p49,Chen:2005p129,Rarity:2005p133,Fulconis:2007p130}.
In TWM, a pump photon
is split into a  pair of photons, while in FWM, two photons in
the pumping beam are
converted into a pair of time-correlated photons.
Therefore, in FWM,  the resolution of two-photon interference 
is twice that of one-photon interference produced by the pumping
light.

Optical fibers can be used for FWM
because of the single-transverse-mode, low loss, propagation in the small 
cores.
High-visibility interference requires a single-mode character that
allows us to introduce the generated photon pairs into fiber-based
networks easily.
High nonlinearity can be achieved 
by confining the optical field to small area.
Therefore, photonic crystal fibers (PCFs) with small core diameters
are expected to be useful.
The phase matching condition for FWM requires that
the pump wavelength be close to the zero-dispersion wavelength of 
the fiber.
Various types of PCFs
with zero-dispersion wavelengths different from
those of conventional fibers
are commercially available.

In this paper, first, we discuss a method for the generation
of photon pairs with a PCF.
We derive the phase matching condition,
which is a function of the dispersion relation 
of a $\chi^{(3)}$ medium,
and obtain the photon pair spectrum
from the group velocity dispersion (GVD)
in the datasheet of the PCF used in our experiment.
The calculated result shows that the width of the photon pair spectrum
is sufficiently high to evade the effect of Raman-scattered photons
when the wavelength of pump light is slightly above the 
zero-dispersion wavelength.
In accordance with the result,
we demonstrate the generation of photon pairs
and obtain $660\,\U{nm}$ / $900\,\U{nm}$ photon pairs 
at a rate of $2,000\ {\rm counts/s}$ and a pump power of $4\,\U{mW}$.
Next, we introduce these photon pairs into
an unbalanced Mach-Zehnder interferometer
with a long arm and a short arm
to verify their time correlation.
In this interferometer, which shares the same principle 
as Franson interferometer \cite{Franson:1989p152,Kwiat:1993},
two quantum states corresponding to the case where
both photons transverse the same arm
contribute to the two-photon interference.
By precise coincidence counting for eliminating the possibility
that each photon follows a different path,
we obtain definite two-photon interference fringes with
83\% visibility exceeding the classical limit,
which provides clear evidence for the time correlation of the photon pairs.

\section{Generation of photon pairs}
\label{generation}

\begin{figure}[tb]
 \begin{center}
  \includegraphics[]{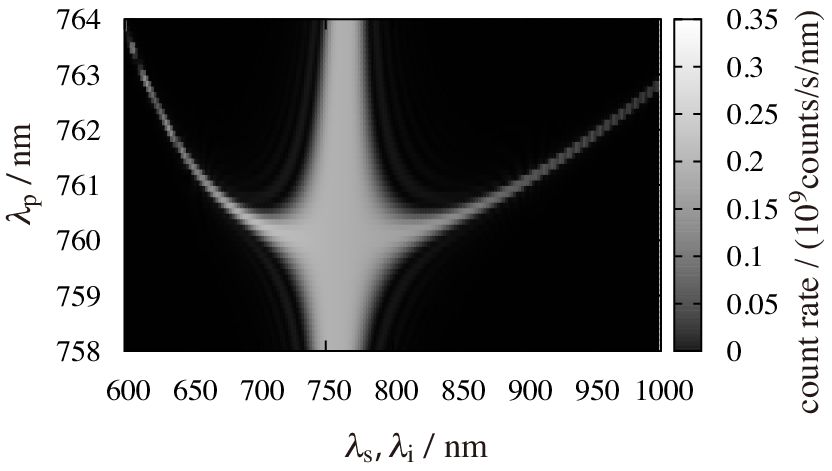}
  \caption{Calculated spectrum of the photon pair ($\lambda\sub{s},
  \lambda\sub{i}$)
  as a function of the pump wavelength $\lambda\sub{p}$. $P=100\,\U{mW}$. }
  \label{Fig:spectrum}
 \end{center}
 \begin{center}
  \includegraphics[]{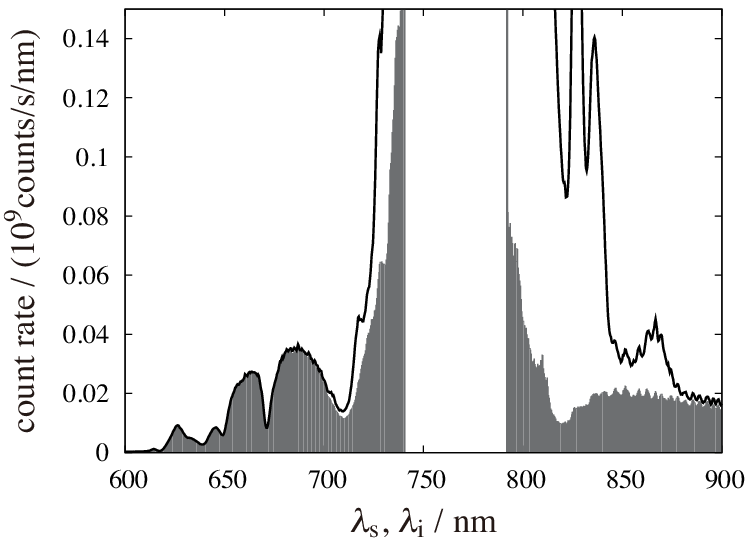} 
  \caption{The solid line represents the output spectrum from the PCF
  for $\lambda\sub{p}=760.4\,\U{nm}$ and $P=100\,\U{mW}$.
  (The spectrum from 740\,nm to 790\,nm is not displayed.)
  The gray region represents the estimated spectrum of the photon pairs.}
  \label{Fig:spectrum-exp}
 \end{center}
\end{figure}

In $\chi^{(3)}$ media, 
two pump photons are converted into a pair of photons,
where one of them is called a signal photon and
the other is called an idler photon, 
through  spontaneous four-wave mixing.
The energy conservation law requires that
\begin{align}
 2 \omega\sub{p} = \omega\sub{s} + \omega\sub{i} , \label{energy_conservation}
\end{align}
where $\omega\sub{p}$, $\omega\sub{s}$, and $\omega\sub{i}$ correspond to
the angular frequencies of the pump, signal, and idler photons, respectively.
The spectrum of  photon pairs is determined by the dispersion relation
$k(\omega)$ of the medium.
The number of  photon pairs within a spectral width $\Delta \Omega$
and a time interval $\Delta t$ is expressed as follows:
\begin{align}
 N &= (\gamma P L)^2 \left|
 \frac{\sin(\kappa L)}{\kappa L}
 \right|^2 \Delta \Omega \Delta t, \label{N}
\end{align}
with
\begin{align}
 \kappa &= \sqrt{\frac{\Delta k}{2} 
 \left(
 \frac{\Delta k}{2} + 2 \gamma P
 \right) 
 } \label{Eq:phasematch},\\
 \Delta k &= k(\omega\sub{s}) + k(\omega\sub{i}) - 2 k (\omega\sub{p}),
 \label{Eq:Deltak}
\end{align}
where $\gamma$ is a nonlinear coefficient, $P$ is the pump power, 
and $L$ is the interaction length \cite{Wang:2001p156}.

Equation (\ref{energy_conservation}) shows that
$\omega\sub{s}$ and $\omega\sub{i}$ are symmetrically located 
with respect to $\omega\sub{p}$.
The spectral separation between the pump photon  and
the photon pairs can be written as $\omega\sub{s}=\omega\sub{p} + \Delta \omega$
and $\omega\sub{i}=\omega\sub{p} - \Delta \omega$.
The separation $\Delta \omega$ should be large in order to evade
spontaneous Raman scattering,
whose spectrum spreads over several hundred nanometers around the pump 
wavelength.

We use a polarization-maintaining PCF
with the zero-dispersion wavelength $\lambda_0=760\,\U{nm}$ 
(Crystal Fiber, NL-PM-760),
$\gamma=102 / \,\U{W / km}$, and $L=1.93\,\U{m}$.
The dispersion relation $k(\omega)$ can be derived from
the group velocity dispersion $D(\omega)$, which is provided by
the manufacturer.
We can derive the spectrum of the photon pair
($\lambda\sub{s}$, $\lambda\sub{i}$)
from Eq.~(\ref{N})
as a function of the pump wavelength $\lambda\sub{p}$,
as shown in Fig.~\ref{Fig:spectrum}.
The spectrum is characterized by two distinct parts:
the trunk near $\lambda\sub{p}$
and the branches bifurcating at $\lambda_0$.
These parts are classified with the phase matching condition $\kappa
\sim 0$ defined in Eq.~(\ref{Eq:phasematch}):
$\Delta k/2+2 \gamma P \sim 0$ for the trunk part and
$\Delta k \sim 0$ for the branch part.
Because the trunk part usually overlaps the spectrum of the Raman-scattered
photons, it is difficult to single out the photon pairs.
In contrast, the spectrum of the branch part
is well separated from that of the Raman-scattered photons
for $\lambda\sub{p} \gtrsim \lambda_0$.

In order to avoid  Raman scattering,
we set the pump wavelength slightly higher than the zero-dispersion
wavelength $\lambda_0 \sim 760\,\U{nm}$.
Figure {\ref{Fig:spectrum-exp}} (solid line) shows
an example of an output spectrum from the PCF
for $\lambda\sub{p}=760.4\,\U{nm}$ and $P=100\,\U{mW}$.
The pump beam from
a CW Ti:sapphire laser was coupled into the PCF,
and the output spectrum was monitored by a spectrometer
after passing it through a notch filter
twice to eliminate the strong pump field.
(The spectrum from $740$ to $790\,\U{nm}$,
which corresponds to the stop band of the notch filter, is not shown
in Fig.~{\ref{Fig:spectrum-exp}}.)
The output spectrum (solid line) shows not only
the photon pairs but also the Raman-scattered photons
and residual pump photons.
It is possible to estimate the fraction of the photon pairs
from the fact that 
the generation rate of the photon pairs is proportional to $P^2$,
while that of the other photons is proportional to $P$.
The gray region in Fig.~{\ref{Fig:spectrum-exp}} represents 
the estimated spectrum of the photon pairs.
The result shows that although the Raman-scattered photons are distributed
widely, the photon pairs dominate the spectrum below $720\,\U{nm}$
and above $880\,\U{nm}$.
We use $\lambda\sub{s}=660\,\U{nm}$ and $\lambda\sub{i}=900\,\U{nm}$
for the  experiments on two-photon interference.

\section{Two-photon interference of photon pairs from PCF}
\subsection{Two-photon interference in unbalanced Mach-Zehnder interferometer}

\begin{figure}[tb]
 \begin{center}
  \includegraphics[width=8.2cm]{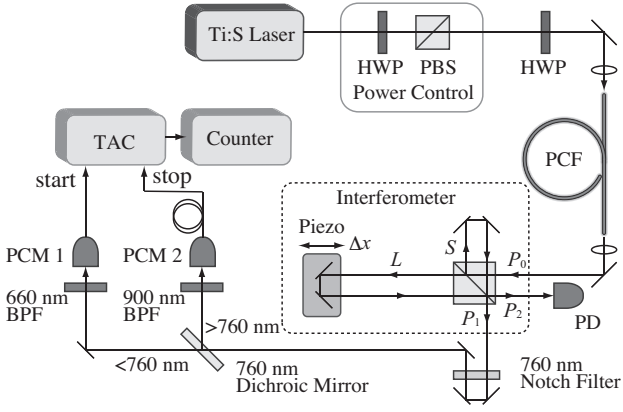}
  \caption{Experimental setup for two-photon interference.
HWP, half-wave plate; PBS, polarizing beam splitter;
BPF, band-pass filter; PCM, photon counting module;
TAC, time-to-amplitude converter.}
  \label{Fig:exp}
 \end{center}
\end{figure}

For the experiment on two-photon interference,
we use an unbalanced Mach-Zehnder interferometer comprising
a long arm ``L'' with a length of $L\sub{L}$ and 
a short arm ``S'' with a length of $L\sub{S}$
as  shown  in Fig.~\ref{Fig:exp}.
The photon pairs are fed to an input port $P_0$, and
they exit from one of the two output port 
$P_1$ and $P_2$, after taking the long path or the short path.

We represent the photon state $\ket{k_a, P_j}$ 
with the wavenumber $k_a \, (a={\rm s, i})$
located at port $P_j (j=0, 1, 2)$.
We calculate the state $\ket{\Psi_1}$ at which we observe 
the signal and idler photons exiting from the same output port $P_1$:
\begin{align}
 \ket{\Psi_1} = \iint \dd k\sub{s} \dd k\sub{i} \Psi(k\sub{s}, k\sub{i})
 \ket{k\sub{s}, P_1} \ket{k\sub{i}, P_1}, \label{psi1}
\end{align}
where 
the spectrum $\Psi(k\sub{s}, k\sub{i})$  is determined by
the phase matching condition and the band-pass filters.
The state $\ket{k_a, P_1}$ is a superposition of 
states in which the photon from $P_0$
follows either the short path $S$ or the long path $L$:
\begin{align}
 \ket{k_a, P_1} = \frac{1}{\sqrt{2}} \ee^{\ii k_a L\sub{L}} \ket{k_a, P_0} 
 + \frac{1}{\sqrt{2}} \ee^{\ii k_a L\sub{S}} \ket{k_a, P_0} \label{Fa}.
\end{align}
Substituting Eq.~(\ref{Fa}) into Eq.~(\ref{psi1}),
we obtain
\begin{align}
 \ket{\Psi_1} &=  \frac{1}{2} \iint \dd k\sub{s} \dd k\sub{i} \Psi(k\sub{s}, k\sub{i}) \,
 \ee^{\ii 2 k\sub{p} L_S} \nonumber \\
  &(
 \ee^{\ii 2 k\sub{p} \Delta L} 
 + \ee^{\ii k\sub{s} \Delta L}
 + \ee^{\ii k\sub{i} \Delta L}
 + 1
 ) \ket{k\sub{s}, P_0} \ket{k\sub{i}, P_0}, \label{psiF}
\end{align}
where we use the relation $k\sub{s}+k\sub{i}=2k\sub{p}$
($k\sub{p}=\omega\sub{p}/c$) derived from Eq.~(\ref{energy_conservation})
and define $\Delta L = L\sub{L} - L\sub{S}$.

If the path difference $\Delta L$ is larger than the coherent length
of each photon,
the probability of finding both the photons at port $P_1$ 
is given as follows:
\begin{align}
  \bracketi{\Psi_1}{\Psi_1} \propto 1 + \frac{1}{2} \cos (2 k\sub{p}
 \Delta L). \label{long}
\end{align}
The visibility of  two-photon interference is only 50\%,
which is equal to the classical limit \cite{Sanaka:2001p115},
because the second and third terms in Eq.~(\ref{psiF}) do not contribute
to it.
These terms correspond to the states where 
each photon follows a different path.
Therefore, it is possible to eliminate them
by counting only the photon pairs within a time interval shorter 
than the propagation time difference.
In other words, we can postselect the first and fourth terms
as follows:
\begin{align}
 \ket{\Psi_1^\prime} = & \frac{1}{2} \iint \dd k\sub{s} \dd k\sub{i}
 \Psi(k\sub{s}, k\sub{i}) \,
 \ee^{\ii 2 k\sub{p} L_S} \nonumber \\
  &(
 \ee^{\ii 2 k\sub{p} \Delta L} 
 + 1
 ) \ket{k\sub{s}, P_0} \ket{k\sub{i}, P_0}. \label{psiFp}
\end{align}
Their probability is
\begin{align}
  \bracketi{\Psi_1^\prime}{\Psi_1^\prime} \propto 1 + \cos (2 k\sub{p}
 \Delta L). \label{short}
\end{align}
All the terms in Eq.~(\ref{psiFp}) contribute to the interference
and produce the interference fringes with 100\% visibility,
which cannot be explained by the classical theory.

\subsection{Experiments}

\begin{figure}[tb]
 \begin{center}
  \includegraphics[]{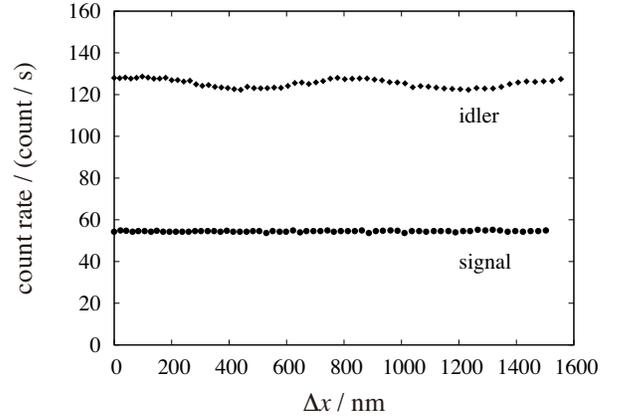}
  \caption{One-photon interference of each photon. Circles (diamonds)
  represent the count rate of the signal (idler) photons as a function
  of $\Delta x$.}
  \label{Fig:one}
 \end{center}
\end{figure}

A schematic representation 
of the experimental setup is shown in Fig.~\ref{Fig:exp}.
The photon pairs from the PCF are sent to the unbalanced 
Mach-Zehnder interferometer.
For elimination of the pumping light,
the output light from $P_1$ is
separated into two paths by a dichroic mirror after 
passing through a notch filter twice.
The signal photon and the idler photon are detected 
by the respective photon counting modules,
whose quantum efficiencies are $32\%$ at $660\,\U{nm}$ (PCM 1)
and $33\%$ at $900\,\U{nm}$ (PCM 2).
The band-pass filters with an FWHM of $10\,\U{nm}$ passband 
are placed in front  of the detectors.
We use a time-to-amplitude converter (TAC) and an electronic counter
for coincidence counting.
The counter counts the number of photon pairs 
within a specific time interval $T$, which
can be controlled by adjusting the discrimination levels of the counter.

\begin{figure}[t]
 \begin{center}
  \includegraphics[]{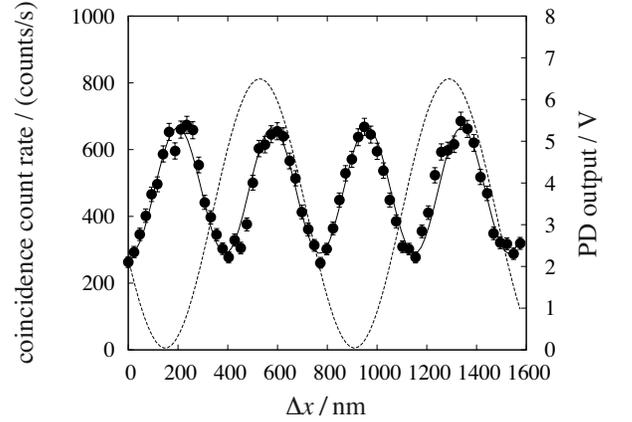}
  \caption{
 The circles represent the coincidence count rates
 for $T=6\,\U{ns}$.
 The one-photon interference pattern of the pump light,
 which is detected by the PD,
 is superimposed for reference (dotted line).}
  \label{Fig:result-long}
\end{center}
\end{figure}
\begin{figure}[t]
\begin{center}
  \includegraphics[]{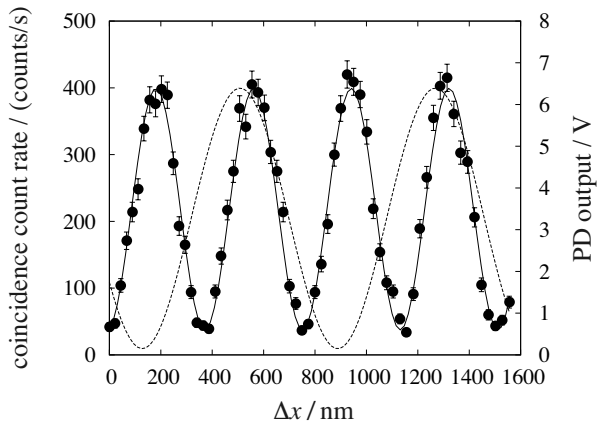}
  \caption{
  The circles represent the coincidence count rates
 for $T=1.5\,\U{ns}$.
  The one-photon interference pattern of the pump light is
 superimposed for reference
 (dotted line).}
  \label{Fig:result-short}
 \end{center}
\end{figure}

We set the path difference $\Delta L \sim 60\,\U{cm}$,
which corresponds to the optical delay of $\tau=2\,\U{ns}$.
The piezo actuator attached to one of the mirrors in the
interferometer provides a variation of $\Delta x$ in $\Delta L$.
We can derive the relation between  $\Delta x$ and the voltage 
applied to the piezo actuator 
from the one-photon interference pattern $\sin(k\sub{p}
\Delta x)$ for the pump field 
by monitoring the signal of the photodiode (PD).
In all experiments, the pump laser is operated at a power of 
$4\,\U{mW}$,
which produces  about $2,000$ pairs per second at the detectors
without the interferometer.
The rate at which photon pairs are produced
is sufficient for the experiment on
two-photon interference, and the accidental coincidence rates 
are negligibly small.

We first measured one-photon interference for the signal and idler photons
by individually counting the photons.
Figure \ref{Fig:one} shows the count rates
as a function of  $\Delta x$.
As expected, there is no clear interference fringe,
because $\Delta L$ is much larger than the coherent length of each
photon.
(The faint fringe for the idler photons can be attributed to
the influence of the residual pump light.)

We performed two types of two-photon interference experiments
by changing the gate time $T$ for coincidence counting.
First, we set $T=6\,\U{ns}\, (> \tau)$ to measure the coincidence
count rate that involves all the photon pairs passing 
through the interferometer.
Figure \ref{Fig:result-long} shows
the one-photon interference pattern
of the pump light (dotted line) for a reference
and the two-photon interference pattern of the photon pairs
(circles).
The period of the two-photon interference fringe
is half that of the one-photon interference fringe for the pump light.
The visibility is only 40\%,
which is consistent with Eq.~(\ref{long}).
Next, we set $T=1.5\,\U{ns}\, (< \tau)$
to eliminate the possibility of 
the signal and idler photons 
following different paths.
The results are shown in Fig.~\ref{Fig:result-short}.
We observed a two-photon interference fringe
with 83\% visibility, which exceeds the classical limit of 50\%.
The result shows that the photon pairs generated in the PCF
have time correlations
that can be explained only by the quantum theory.
The main reason for the degradation of the visibility is mainly
due to the chromatic aberration of the objective lens
that collimates the photon pairs from the PCF.
By using an objective lens specially designed for red 
and infrared wavelengths, the visibility of two-photon 
interference will be improved.

\section{Conclusion}

In this paper, we report the generation of photon pairs by
four-wave mixing in the photonic crystal fiber and 
the two-photon interference of the photon pairs.
By maintaining the wavelength of the pump light
slightly above the zero-dispersion
wavelength, we obtained a wider spectrum of photon pairs 
as compared to that of Raman-scattered photons. 
We set the pump light wavelength to $760.4\,\U{nm}$ and
obtained $660\,\U{nm}$/$900\,\U{nm}$
 photon pairs at $2,000\,{\rm count/s}$
with a pump power of only $4\,\U{mW}$,
which could be generated by a diode laser.
Because the photon pairs were emitted in a 
single-transverse mode of the fiber,
all of them contributed to the interference effectively
without any spatial filtering.
The high brightness of the photon pairs per mode
is an advantage of the fiber-based sources 
over crystal-based sources.

For higher photon-pair flux, we can use an intense pulse laser
with slight modification in the two-photon interferometer \cite{Brendel:1999}.
In general, the conversion efficiency increases with the fiber length.
However, due to the losses, a fiber longer than the fiber attenuation
length, which is about $86\,\U{m}$ for our fiber with $50\,\U{dB/km}$
losses, is of no use.

Two-photon interference achieved using the unbalanced Mach-Zehnder 
interferometer
produced clear interference fringes with 83\% visibility,
which exceeds the classical limit.
The period of the fringes was half that of the pump light wavelength,
which is  different from the case of two-photon interference
where photon pairs were generated through TWM.
This shows that two-photon interference of photon pairs through FWM
can be beneficial to high-resolution lithographic technology 
and phase-sensitive interferometries.

\section*{Acknowledgements}

This research is supported by
the global COE program ``Photonics and Electronics Science and 
Engineering,'' 
at Kyoto University and
a Grant-in-Aid for Young Scientists (B), No.~18740247,
from the Japan Society for the Promotion of Science.

~

\end{document}